\documentclass{svjour3}                     
\smartqed  
\usepackage{graphicx}
\usepackage{natbib}
\bibpunct{(}{)}{;}{a}{}{,}

\newcommand{\apj}{ApJ}
\newcommand{\apjl}{ApJL}

\newcommand{\nat}{Nature}
\newcommand{\aap}{A\&A}

\newcommand{\pasj}{PASJ}

\newcommand{\araa}{ARA\&A}
\newcommand{\solphys}{Sol. Phys.}

\begin{document}

\title{Recent Evidence for Convection in Sunspot Penumbrae 
}


\author{G{\"o}ran B. Scharmer        
}


\institute{G.B. Scharmer \at
              Institute for Solar Physics, 
              Royal Swedish Academy of Sciences,
              AlbaNova University Center,
              SE-10691 Stockholm\\
              \email{scharmer@astro.su.se}           
}

\date{Received: date / Accepted: date}

\maketitle

\begin{abstract}
Whereas penumbral models during the last 15 years have been successful in
explaining Evershed flows and magnetic field inclination variations in terms
of flux tubes, the lack of contact between these models and a convective
process needed to explain the penumbral radiative heat flux has been disturbing.
We report on recent observational and theoretical evidence that challenge
flux tube interpretations and conclude that the origin of penumbral filamentary 
structure is overturning convection.

\keywords{First keyword \and Second keyword \and More}
\end{abstract}

\section{Introduction}
\label{intro}
Sunspot magnetic fields and dynamics have been studied scientifically for 100 years.
Despite considerable progress during the last decade, a theoretical framework 
that explains sunspot fine structure, dynamics, magnetic fields and energy balance 
in a consistent manner is only now beginning to emerge. This situation can partly
be attributed to the small horizontal scales associated with sunspot fine structure
and the relatively poor spatial resolution achieved with spectropolarimetric
observations. In addition, realistic numerical 3D MHD simulations of sunspots have 
only recently become possible. 

During the last few years, there has been a remarkable improvement in the quality 
and diversity of observational data relevant to the understanding of sunspot fine 
structure, dynamics, magnetic fields and energy balance. In particular, high-spatial 
resolution observations from the Swedish 1-m Solar Telescope (SST) and the Solar Optical 
Telescope (SOT) on Hinode reveal new sunspot structure and flow patterns at odds with 
prevailing interpretations in terms of flux tube models. In addition, theoretical 
arguments as well as recent 3D MHD simulations of sunspot fine structure underline 
problems of these interpretations and lead to the conclusion that the origin of 
penumbral fine structure is overturning convection. 

In the present review, we describe recent progress in our understanding of
penumbral fine structure and put that in context with existing models. Rather than
attempting to summarize the extensive literature on penumbrae, we discuss selected
key papers and attempt to describe their interconnections and to critically review 
conclusions drawn. We also point out
connections between observed penumbral fine structure and magnetic flux concentrations 
{\em outside} sunspots, such as faculae. We hope to convince the reader that the 
new picture emerging is one of improved consistency as regards observations and 
theory of sunspot penumbrae in particular, but also with respect to umbral dots, light 
bridges and faculae.

\section{Overview of Established Models}
\label{Models}

For the past 15 years, the predominant paradigm of penumbral filaments has been based on 
nearly horizontal flux
tubes, portrayed and modeled as radially aligned cylinders, embedded in a more 
vertical magnetic field. These flux tube interpretations have their roots in the work 
of \citet{1968MitAG..25..194M}
who proposed to explain Evershed flows as siphon flows, 
originating from a difference in magnetic field strength between the two foot points 
of a flux tube. In the 70's, flux tubes and clusters of flux tubes were also established
in models of magnetic flux concentrations, surrounded by field-free gas, 
with scales ranging from less than 100~km to that of a large sunspot.    

\subsection{Embedded flux tubes}
\label{fluxtubes}

Of particular importance in the current sunspot literature is the {\em uncombed penumbra} 
model, proposed by \citet{1993A&A...275..283S}
to explain the strongly asymmetric Stokes V profiles observed on the
limb side penumbra for sunspots away from disk center. This model addressed an apparent
problem of very strong line-of-sight (LOS) {\em gradients} in the inclination angle of the 
penumbral magnetic field, inferred from Stokes data by \cite{1992ApJ...398..359S}.
The large LOS inclination gradients derived from this data were
(incorrectly, see Sect. 4) interpreted to imply volume currents and associated curvature forces 
strong enough to completely disrupt static force balance in the spot 
\citep{1992ApJ...398..359S,1993A&A...277..639S}.
The uncombed penumbra model avoids 
this problem, at least partly, by {\em postulating} the existence of discrete flux tubes, 
within which the magnetic field is assumed to be homogeneous and therefore current-free. 
The uncombed model thus `replaces' smooth inclination gradients and (assumed) large 
volume currents with discontinuous changes at the boundary of the flux tube and an 
associated current sheet. Solanki and Montavon demonstrated that a nearly horizontal flux tube, 
with a strong flow parallel to its magnetic field, can explain the observed net circular 
polarization (NCP) resulting from Stokes V asymmetries of this configuration. Moreover, 
if such a flux tube is located {\em entirely} above the photosphere, both its upper 
and lower boundaries contribute with the same sign to the asymmetry of its Stokes V 
profile, thus enhancing the NCP.

Various implementations of flux tube models with polarized radiative transfer
were later developed by e.g. \citet{2000A&A...361..734M, 2002A&A...381..668S, 2007ApJ...666L.133B,
2007A&A...471..967B, 2007ApJ...671L..85T, 2004A&A...427..319B, 2003A&A...403L..47B}.
These calculations demonstrate consistency between the
calculated azimuthal variation of NCP and measurements made at low spatial resolution in 
visible and near infrared spectral lines. Two-component inversions interpreted within 
the context of embedded flux tube models and applied to low spatial resolution
(0.6--1 arcsec) Stokes data by \citet{2004A&A...422.1093B, 2005A&A...436..333B,
2006A&A...450..383B}
similarly were shown to be largely consistent with the assumed inversion (flux tube) model.

\subsection{Siphon flow and dynamic flux tube models}
\label{fluxtubes}
The first flux tube model proposed to explain Evershed flows in penumbrae is 
the siphon flow model of \citet{1968MitAG..25..194M}. In this model, a difference
in field strength between the two footpoints of a flux tube leads to a difference
in gas pressure, driving a flow in the direction of the footpoint with the highest
field strength.
This work was later followed up by \citet{1989A&A...222..297D,1991A&A...248..637D}
and \citet{1988ApJ...333..407T}
and the model further refined
in a series of papers by \citet{1989ApJ...337..977M,1993ApJ...402..314M,
1997Natur.390..485M, 1990ApJ...359..550T, 1991ApJ...375..404T}.
Given the free parameters of the calculations, the siphon flow model of 
\citet{1997Natur.390..485M} allow consistency with the discovery that 
the Evershed flow connects to patches of opposite magnetic polarity at deep layers near
the outer boundary of a sunspot \citep{1997Natur.389...47W}.

Siphon flow models allow an interpretation of Evershed flows as steady flows, with the 
properties of the flow determined by assumed conditions at the footpoints of the flux 
tube and of the surrounding atmosphere. The mechanism that {\em produces} the magnetic 
field strength difference between the footpoints, needed to generate a gas pressure 
gradient to drive the flow, is not explained by such models. \citet{1994A&A...290..295J} 
proposed the concept of interchange convection of magnetic flux tubes (or rather, sheets) 
as an explanation of the penumbral heat flux. To investigate this proposal, 
\citet{1998ApJ...493L.121S, 1998A&A...337..897S}
developed a simplified 1D, one-component numerical model of such a flux tube and studied 
its time evolution. In this model, a flux tube initially in contact with the 
magnetopause (the outer boundary of the sunspot) is heated radiatively by the external 
field-free gas. Its subsequent evolution is driven by the buoyancy of the flux tube and
the superadiabatic stratification of the surrounding penumbra atmosphere, assumed to
have properties unaffected by the flux tube. At the surface, radiative cooling of the tube 
causes it to loose buoyancy such that its upper part settles at a height of about 100~km 
above the photosphere. A gas pressure gradient, driving the Evershed flow, develops along 
the tube from downstream radiative cooling.

Later simulations by {\citet{2002AN....323..303S,2003ASPC..286..211S} with reduced 
numerical viscosities show a similar initial behavior of the flux tube. However,
near the surface, the flux tube subsequently develops standing waves downstream from the 
footpoint with downflows diving down into the convectively unstable layers beneath the surface.
The crests of this oscillating flux tube remain visible above the surface and show an
inward migration in the inner part of the umbra and an outward migration in the outer
penumbra and outside the penumbra. This behavior is similar to that of observed penumbral 
grains in the inner and outer penumbra and moving magnetic features outside the penumbra. The
discovery of small-scale bipolar magnetic features propagating from the mid penumbra
to outside the penumbra, where they become moving magnetic features, is consistent with
the  `sea serpent' behavior of Schlichenmaier's moving flux tubes \citep{2008A&A...481L..21S}.

A problem, investigated by several authors, e.g., \citet{1993A&A...277..639S, 2006A&A...454..975R},
is the large radial mass flux of the Evershed flow inside the penumbra. Only part of this
flow appears to continue in the magnetic canopy above the quiet sun photosphere
outside the spot. To explain this, most of the Evershed flow must submerge close to the outer
boundary of the penumbra. The moving tube model simulations show
such downflows within the penumbra. However, \citet{2005A&A...440L..29T,2006ASPC..354..224T} 
objected that the
undulations seen in the `sea serpents' of \citet{2002AN....323..303S} should occur
preferentially in the {\em horizontal} plane and hence can explain neither moving grains nor
convective downflows. 

The question of a heating mechanism to explain the penumbral radiative heat flux was investigated
by \citet{2003A&A...411..257S}. Based on estimates of the radiative cooling time and the time
span of successive emergences of flux tubes, they concluded that interchange convection cannot
provide the needed energy flux. \citet{2004ApJ...600.1073W} also argued against interchange
convection on the basis that long loops of magnetic field connecting to a distant active region
cannot possibly interchange with horizontal fields carrying Evershed flows. 

The conclusion of \citet{2003A&A...411..257S} was that upflows along the magnetic flux tubes
can explain the penumbral brightness, but only if the flux tube submerges again within a
distance of 1000--2000~km from their footpoint. The upflow in a narrow tube cannot supply the 
radiative energy losses over a distance corresponding to 
the entire radial extent of a penumbra unless it submerges and is re-heated. Such 
re-heating does not solve the energy flux problem, however, since it relies on a (convective)
mechanism to transport the heat to the bottom of the flux tube. Nevertheless, the discovery of
field lines returning to the penumbra and associated downflows \citep{1997Natur.389...47W} 
was considered as support for this explanation \citep{2003A&A...411..257S}.

\subsection{Convection and downward pumping of magnetic flux}
\label{Pumping}

The siphon flow model of Montesinos and Thomas is unrelated to any convection process
operating in the penumbra. This model represents a stationary solution that cannot explain 
time dependent behavior such as moving penumbral grains \citep{2006ASPC..354..224T}.
These grains are instead interpreted as originating from a moving convective pattern
in the brighter parts of the penumbra \citep{2002AN....323..371W, 2006SSRv..124...13W,
2006ASPC..354..213W}. Whereas the moving tube simulations show localized downflows inside
and outside the penumbra, the arched flux tubes of the siphon flow model
require a mechanism to submerge and hold down their outer parts to sustain equilibrium
\citep{1997Natur.390..485M}. 
\citet{2002Natur.420..390T} and \citet{2004ApJ...600.1073W} proposed that this submergence of 
the flux tubes occurs as the result of downward pumping by convection {\em outside} the 
sunspot. They even took this proposal one step further and proposed that this downward 
pumping is the {\em origin} of the filamentary structure of the penumbra. 
In this view, the salient features of penumbrae: their filamentary structures, the strong
variations in magnetic field inclination across filaments and the Evershed flows, are to a
large extent explained by what happens {\em outside} the sunspot. Magnetic fields in bright 
and dark filaments are distinct and cannot be interchanged \citep{2004ARA&A..42..517T}. To 
support this, \citet{2004ARA&A..42..517T, 2004ApJ...600.1073W, 2006ASPC..354..213W} refer to X-ray 
observations and TRACE images showing loops extending over great distances across the Sun.
We believe that their description and connection to the interlocking comb structure of the
penumbra is misleading. Virtually all information about large fluctuations in the 
magnetic field inclination
within the penumbra comes from spectral lines formed within a few hundred km above the photosphere.
The images referred to \citep{1992ApJ...399..313S} do not have the spatial resolution needed 
to separate X-ray loops (interpreted to outline field lines) from bright and dark filaments.
As far as the author knows, there are no observations that allow us to conclude that azimuthal
variations in the magnetic field inclination associated with filamentary structures
correspond to field lines that are widely separated also far away from the penumbral photosphere.
We have argued \citep{2006A&A...447..343S} that the strong inferred variations in the magnetic field
inclination within the first one or two hundred km above the penumbral photosphere, seen even in the
inner penumbra, cannot be explained by a mechanism operating outside the sunspot. Instead,
these strong variations suggest a {\em local} mechanism at work. The bright
and dark filaments are not distinct and they can interchange. Furthermore, Evershed flows are
associated with field lines that only locally and during a limited time are nearly horizontal.

\section{Limitations and problems of flux tube interpretations}
\label{Objections}
The success of the uncombed penumbra model \citep{1993A&A...275..283S} in explaining 
observed Stokes spectra and net circular polarization (NCP) \citep{2007ApJ...666L.133B,
2007A&A...471..967B, 2007ApJ...671L..85T, 2004A&A...427..319B, 2003A&A...403L..47B} 
is unquestionable. Furthermore, the moving tube 
simulations of \citet{2002AN....323..303S} make excellent contact with the uncombed penumbra 
model. It demonstrates consistency with observed strong upflows in bright grains 
\citep{2006ApJ...646..593R} and the behavior of bright grains and moving magnetic features 
\citep{2008A&A...481L..21S}. It is hardly surprising that the ability of these models to explain 
azimuthal and line-of-sight (LOS) gradients of the penumbral magnetic field and Evershed flows 
was deemed successful. 

However, a fundamental problem remains.
Whereas the embedded flux tubes simulated by Schlichenmaier are consistent with 
observations in many respects, such flux tubes present problems in explaining penumbral 
heating \citep{2006A&A...447..343S,2006A&A...460..605S}. As discussed above, a horizontal flux 
tube is likely to heat the penumbra over a radial distance not much more than 1000~km 
\citep{1993A&A...275..283S, 2003ASPC..286..211S}, which is typical of a
penumbral grain rather than a penumbral filament. Even over such a short distance, radiative
cooling of the flow leads to significant temperature and brightness gradients along the 
flux tube unless there is a separate source of heating below the flow channel 
\citep{1999A&A...349..961S}. Spruit and Scharmer therefore argued that the presence of flux tubes 
covering a large fraction of the penumbral surface would constitute a {\em hindrance} for heating 
of the penumbra. They also pointed out that the existence of {\em elevated} flux tubes extending 
up to a few hundred km above the penumbral photosphere
correspond to unlikely perturbations in a magnetic field so dominant already at this height 
that it must be expected to be nearly potential. Furthermore, the moving tube simulations 
represent a highly idealized model that cannot not be expected to be more than a coarse 
representation of reality. In particular:
\begin{itemize}
\item The existence of the flux tube is an {\em assumption} in the model
\item The model is 1-dimensional and the flux tube assumed to be `thin' (see below)
\item The simulations correspond to a 1-component model with the properties of the {\em background} 
      atmosphere unaffected by the evolution of the flux tube
\item Only a single flux tube is simulated. The influence of neighboring flux tubes is not 
      accounted for.
\item The curvature forces of the surrounding magnetic field are ignored and its influence is
      reduced to a scalar magnetic pressure, similar to a gas pressure.
\end{itemize}
The same objections apply to the siphon flow models discussed in previous sections.

In view of these short-comings, it is remarkable that the moving tube simulations appears to
capture important properties of penumbral dynamics. This is further discussed in Sect\@. 4.4.

Magnetostatic flux tube models including forces from a surrounding potential magnetic field
\citep{2007A&A...471..967B} demonstrate the difficulties of embedded flux tube configurations.
Prescribing a specific (circular) cross section for the flux tube corresponds
to an overconstrained problem such that not only the gas pressure but also the temperature and
density within the flux tube are given by force balance alone. There is thus no room for an 
energy equation with this type of models. Furthermore,
equilibrium is not possible with a purely potential magnetic field inside the flux tube. In
the models shown there is an azimuthal component, corresponding to a volume current aligned 
with the flux tube, in addition to the radial field component. The bottom part of the
flux tube is nearly evacuated whereas the top part is denser than the surroundings
in order to balance the magnetic forces at the top and bottom, stretching and flattening
the flux tube. These problems originate from the surrounding magnetic field wrapping around the
flux tube and cannot be resolved by making the flux tube thinner.

In attempt to understand the temperature structure and energy balance of penumbral flux tubes,
\citet{2008A&A...488..749R} developed a model for a flux tube with a weak magnetic field
aligned with a homogeneous magnetic field along the flux tube axis. An objection against 
this model is that it suffers from a lack of consistency as regards force balance, 
which is implemented in a way that is equivalent to ignoring the vector properties of the 
surrounding magnetic field. The origin of this problem is the same as that
of the models of \citet{2007A&A...471..967B}: Prescribing the shape of the flux tube cross section  
is in general incompatible with either force balance or an energy equation. 
   
\subsection{Ambiguities of interpretations based on inversions} 

A major obstacle to understanding penumbra fine structure has been the lack of adequate spatial
resolution in observed polarized and unpolarized spectra. In spite of successful adaptive optics 
systems operating on major solar telescopes, it has not been possible to reach the diffraction 
limit with the long integration times needed for such data with 
adequate signal-to-noise. (However, by combining many {\em short} exposure frames and using image
restoration techniques, filter-based systems allow near diffraction limited spectropolarimetry 
\citep{2008A&A...489..429V, 2008arXiv0806.1638S}.) 
The exception is observations in the near infrared, for example those
made with the Tenerife Infrared Polarimeter, TIP \citep{1999AGAb...15...89M} on the German Vacuum
Tower Telescope (VTT) at wavelengths around 1.5~$\mu$. At that wavelength, the diffraction limited 
resolution is about 0.6~arc sec with the VTT.

\begin{figure}
\hfill
\includegraphics[bb=14 286 286 536,clip,width=0.49\textwidth]{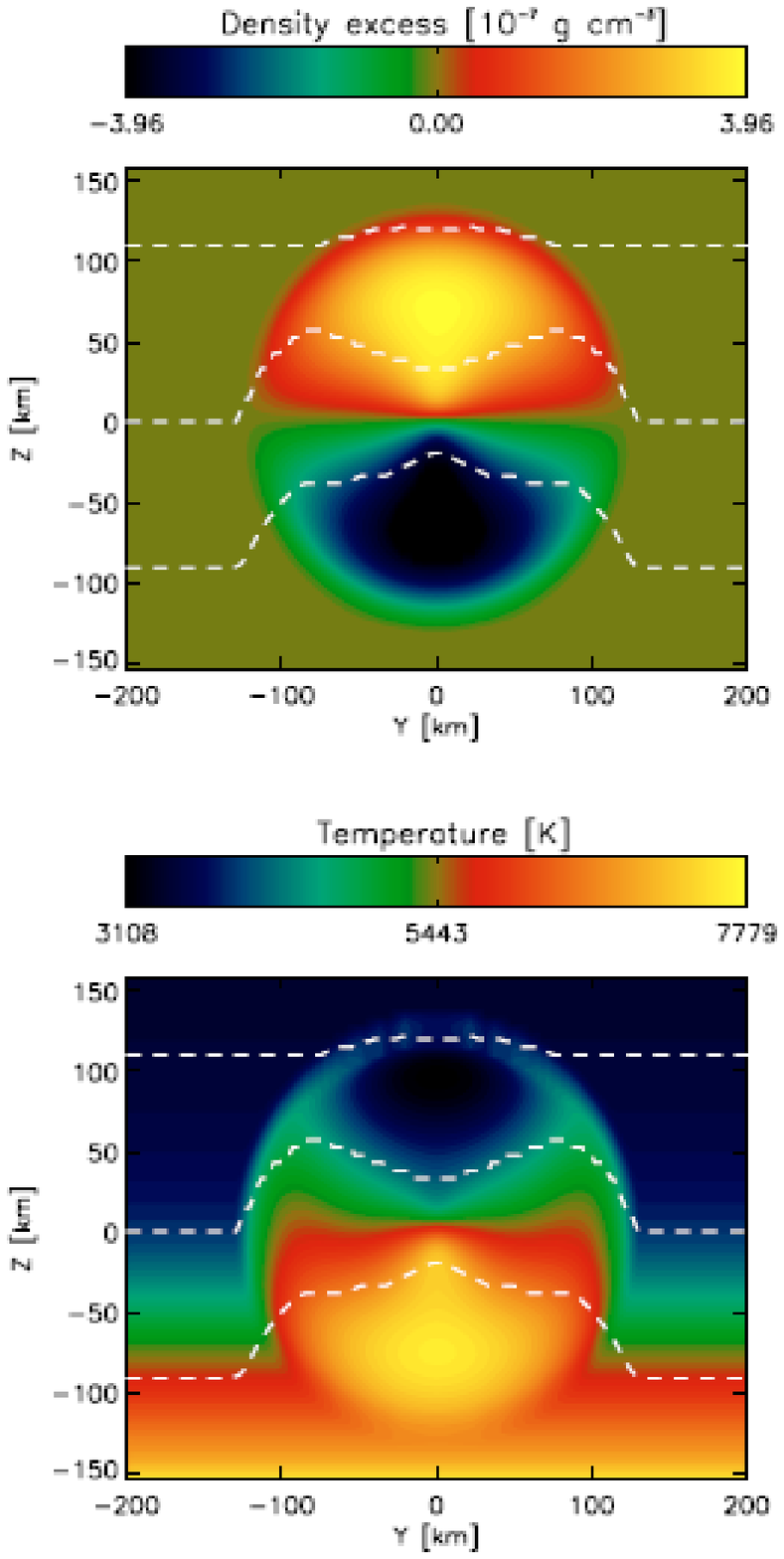}
\hfill
\fbox{\includegraphics[bb=64 70 208 200,clip,width=0.47\textwidth]{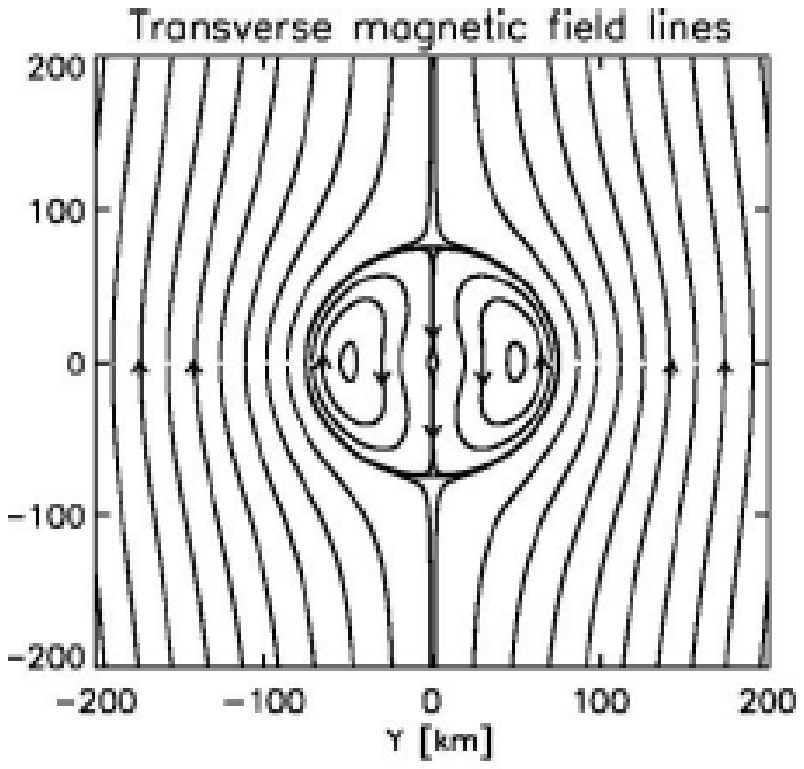}}

\hfill
\caption{\textbf{Left:} Magnetostatic flux tube model of \citet{2007A&A...471..967B} with dashed 
curves showing (top to bottom) optical depths $\tau$ of 0.1,1 and 10. Note that only the upper 
part of the flux tube is located above the photosphere and that nearly the entire flux tube is
located below $\tau=0.1$. \textbf{Right:} Transverse field lines of a similar (but not 
identical) model \citep{2007ApJ...666L.133B}. Note the similarity of the magnetic field 
configuration {\em above} the flux tube in this model and that of the convective gap models, shown
in Fig\@.\ref{gap}}
\label{flux_tube}       
\end{figure}

At a spatial resolution of 0.6 arc sec or worse, penumbral fine structure is not adequately 
resolved. To compensate for this and in order to test the validity of the embedded flux tube 
(uncombed penumbra) model, various observers, e.g., \citep{2004A&A...422.1093B, 2006A&A...450..383B,
2003A&A...403L..47B, 2004A&A...427..319B} have implemented two-component inversion techniques to 
interpret Stokes spectra. These investigations show that it is indeed possible to 
fit the data with the assumed (highly idealized) representations of flux tubes. Forward 
calculations of Stokes spectra \citep{2000A&A...361..734M} and unpolarized spectra 
\citep{1995A&A...298..260R,2006A&A...453.1117B} based on flux tube models also demonstrate 
consistency. In some cases, it was demonstrated also that the observed data could equally well 
be reproduced with flux tube representations and models with smooth gradients 
\citep{2000A&A...361..734M, 2002A&A...389.1020R, 2004A&A...422.1093B, 2006A&A...453.1117B}. 
This ambiguity is a consequence of the width of the radiative transfer response function, smearing 
out the effects of discontinuities in the observed (Stokes) spectra. 
The interpretations of Stokes spectra clearly show compatibility with flux tube interpretations. 
However, the simplicity of the implemented inversion
models and the use of two components to represent observational data of penumbral
fine structure at inadequate spatial resolution adds to these uncertainties to the extent
that we are justified in questioning whether a description in terms of embedded flux tubes is 
an adequate representation of penumbra fine structure. 

\section{Convective origin of penumbral filaments}

An alternative explanation to understanding penumbra fine structure was proposed by 
\citet{2006A&A...447..343S}. 
The filamentary structure is explained by convection in radially aligned (nearly) 
field-free gaps just below the visible surface. Such intrusions unavoidably lead to strong 
variations in the magnetic field strength and inclination above the gaps, but these variations
are fully consistent with even a simple {\em potential} magnetic field configuration. 
The model explains dark cores in bright filaments \citep{2002Natur.420..151S}, seen in the inner
and mid penumbra (c.f., Fig\@.s \ref{cores1} and \ref{cores2}) as an indicator of strong field 
strength variations across filaments, leading to a strongly varying Wilson depression.
The Evershed flow is in this model identical to the horizontal flow 
component of the convection \citep{2008ApJ...677L.149S}. We explain this model in more 
details in the following.  

\begin{figure*}[tb]
\centering
        \includegraphics[width=\textwidth]{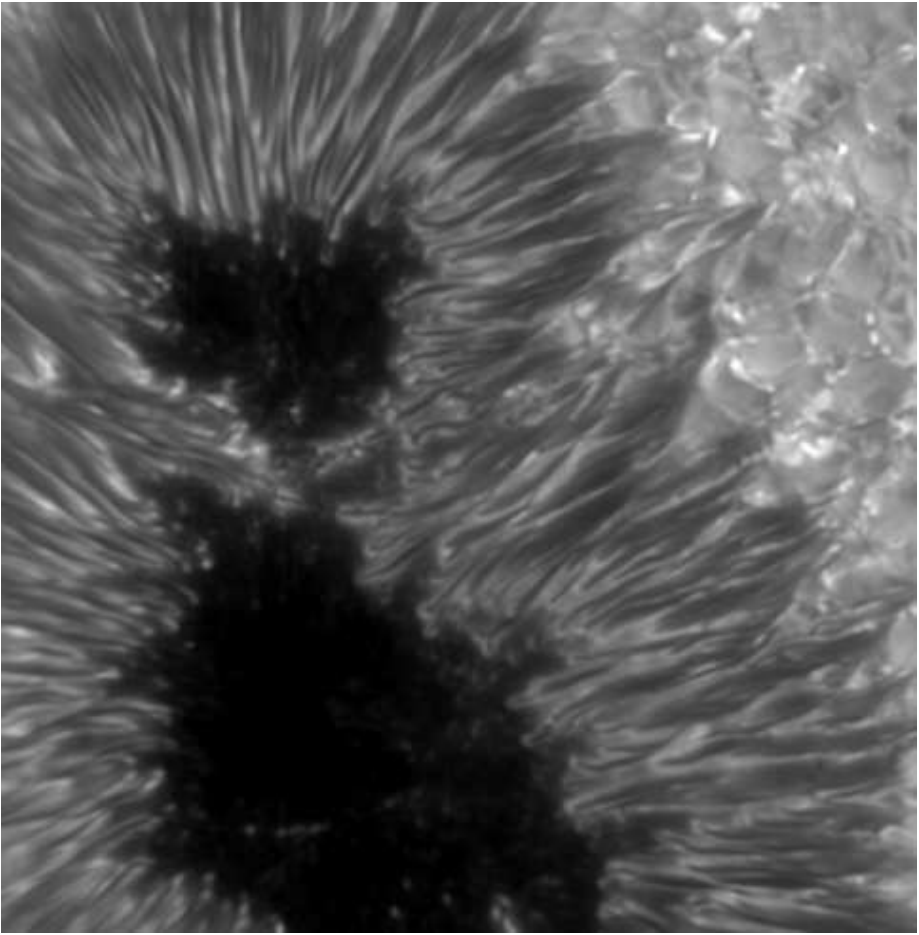}
     \caption{Sunspot located at a heliocentric distance of 20~deg, observed with the SST on 12 
              Sep 2006 \citep{2008A&A...489..429V}. The image shows the Stokes I intensity, averaged 
              over the
              blue and red wings of the 6302 iron line. Note the dark cores, clearly visible in the
              {\em inner} penumbra.
       }
     \label{cores1}
\end{figure*}

\begin{figure*}[tb]
\centering
        \includegraphics[width=\textwidth]{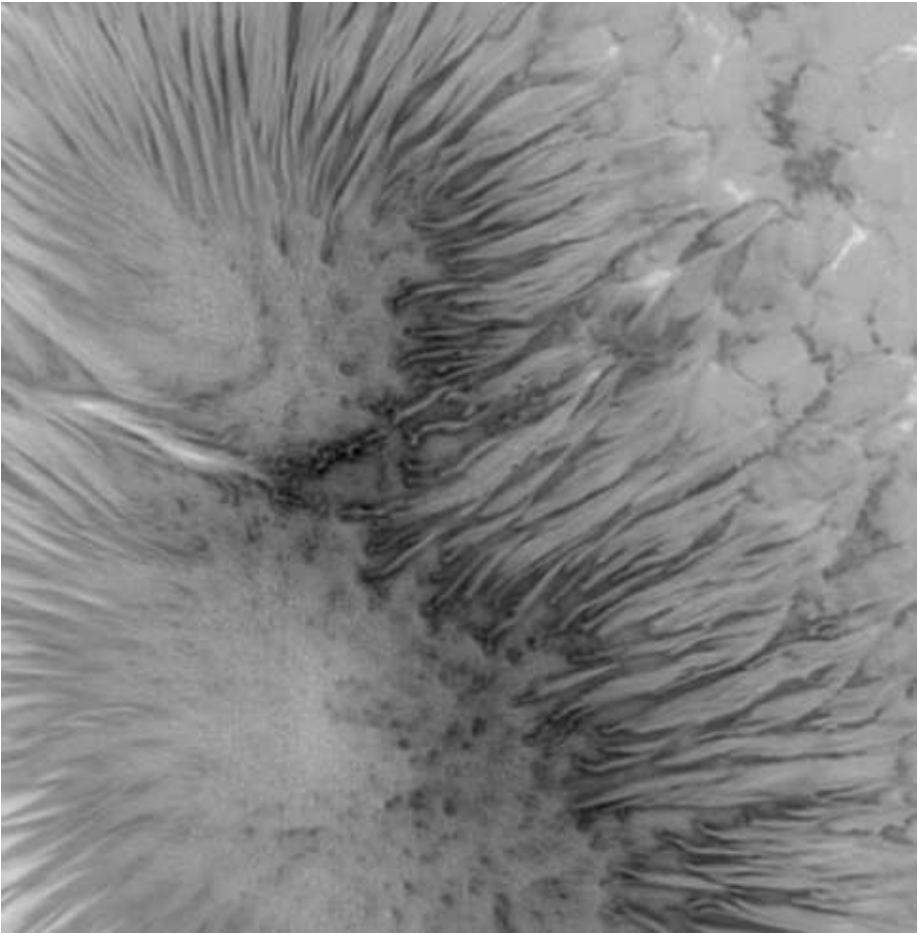}
     \caption{The same sunspot as shown in Fig\@.\ref{cores1} \citep{2008A&A...489..429V}.
              The image shows the difference between
              Stokes V (circularly polarized light), recorded in the blue and red wings of the
              6302 iron line. This serves as a proxy for the line-of-sight component of the
              magnetic field vector. Note the strongly reduced polarization signal at locations
              of the dark cores, suggesting strongly reduced field strength at these locations.
       }
     \label{cores2}
\end{figure*}

The large inclination gradients are a consequence of magnetic fields being divergence free 
($\bf\nabla ^{\bf.} \bf{B} = 0$): field lines cannot disappear at the top of a gap (or flux tube)
but must bend around it. To estimate the characteristic vertical scale $H$ for these 
inclination variations, we can assume a potential magnetic field 
\citep{2006A&A...447..343S}. For filaments separated by a distance $L$, this gives 
$H \approx L/2\pi$. With a typical separation of 1" between filaments, this 
corresponds to a vertical height scale of 120~km. Simple magnetostatic models 
for such configurations, based on identical temperature variations with height 
for the two components, show distinct differences between the inner penumbra, where the 
magnetic field is more vertical and stronger than in the outer penumbra 
\citep{2006A&A...460..605S}. In the {\em inner} penumbra, the magnetic field is cusp-shaped 
above the gaps and associated with a large ($\approx$~200--300~km) Wilson depression
relative to that of the gaps. Even when the temperature is the same inside and outside the gaps, 
the strong Wilson depression leads to an observed brightness that is lower above the gaps 
than between the gaps. The {\em dark-cored filaments} discovered with the Swedish 1-m Solar Telescope 
(SST) \citep{2002Natur.420..151S}, seen in the inner and mid penumbra, are thus explained by a 
{\em combination of increased opacity associated with a strongly reduced field strength and
an overall drop of temperature with height} \citep{2006A&A...447..343S}. In the {\em outer} 
penumbra, the Wilson depression is only on the order of 50~km in these models. With such a 
small Wilson depression, intensity variations 
from `global' vertical temperature gradients cannot be expected to completely dominate over 
{\em local} horizontal and vertical gradients associated with details of the heating and 
cooling of convecting gas. In the {\em outer} penumbra we therefore do not expect the same
kind of relation between filament brightness and field strength as for the inner penumbra.
The magnetostatic models are therefore in qualitative agreement with the absence of dark 
cores in the outer penumbra.

\begin{figure}
\hfill
\includegraphics[bb=235 93 590 725, angle=-90,clip, width=0.49\textwidth]{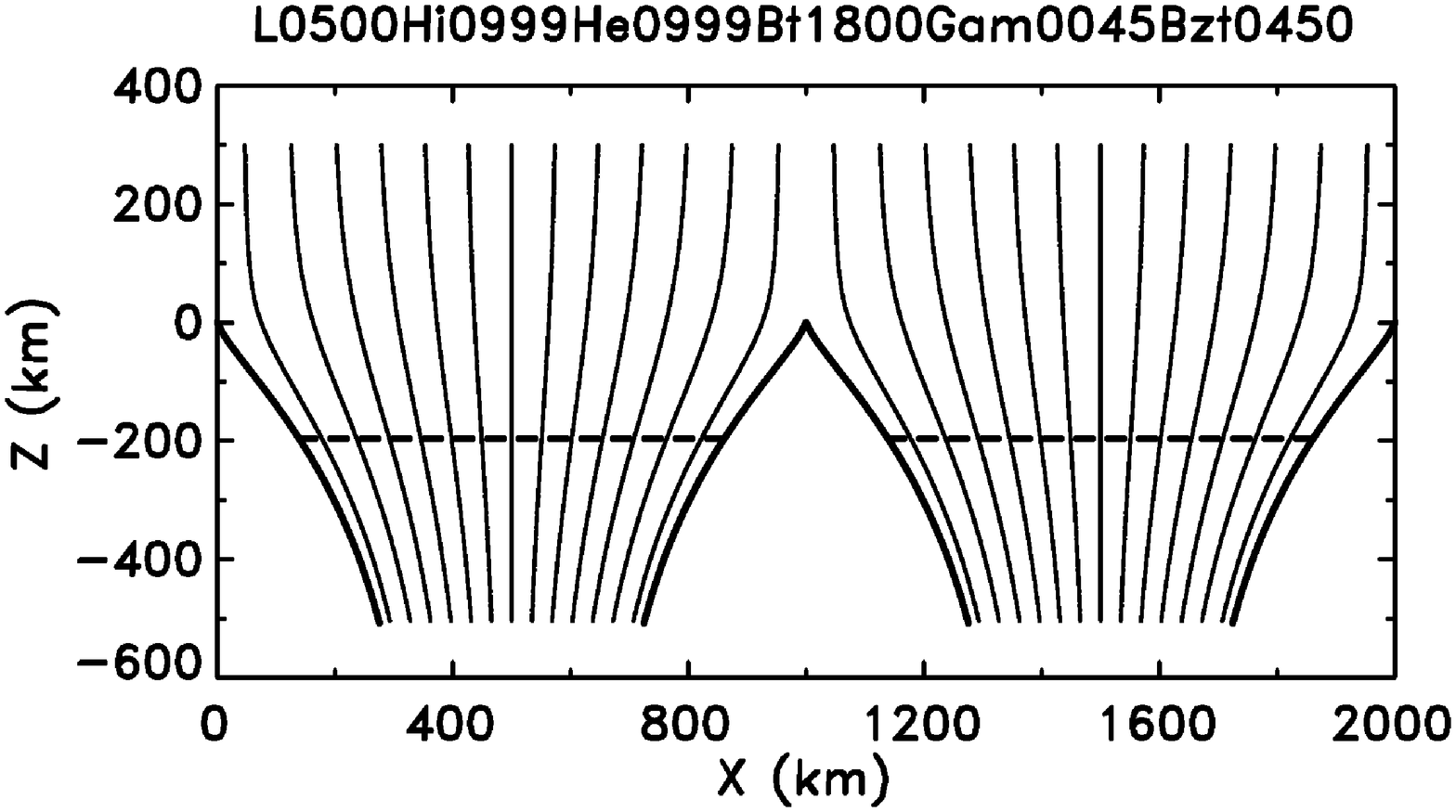}
\hfill
\includegraphics[angle=-90, bb=235 93 590 725, clip, width=0.49\textwidth]{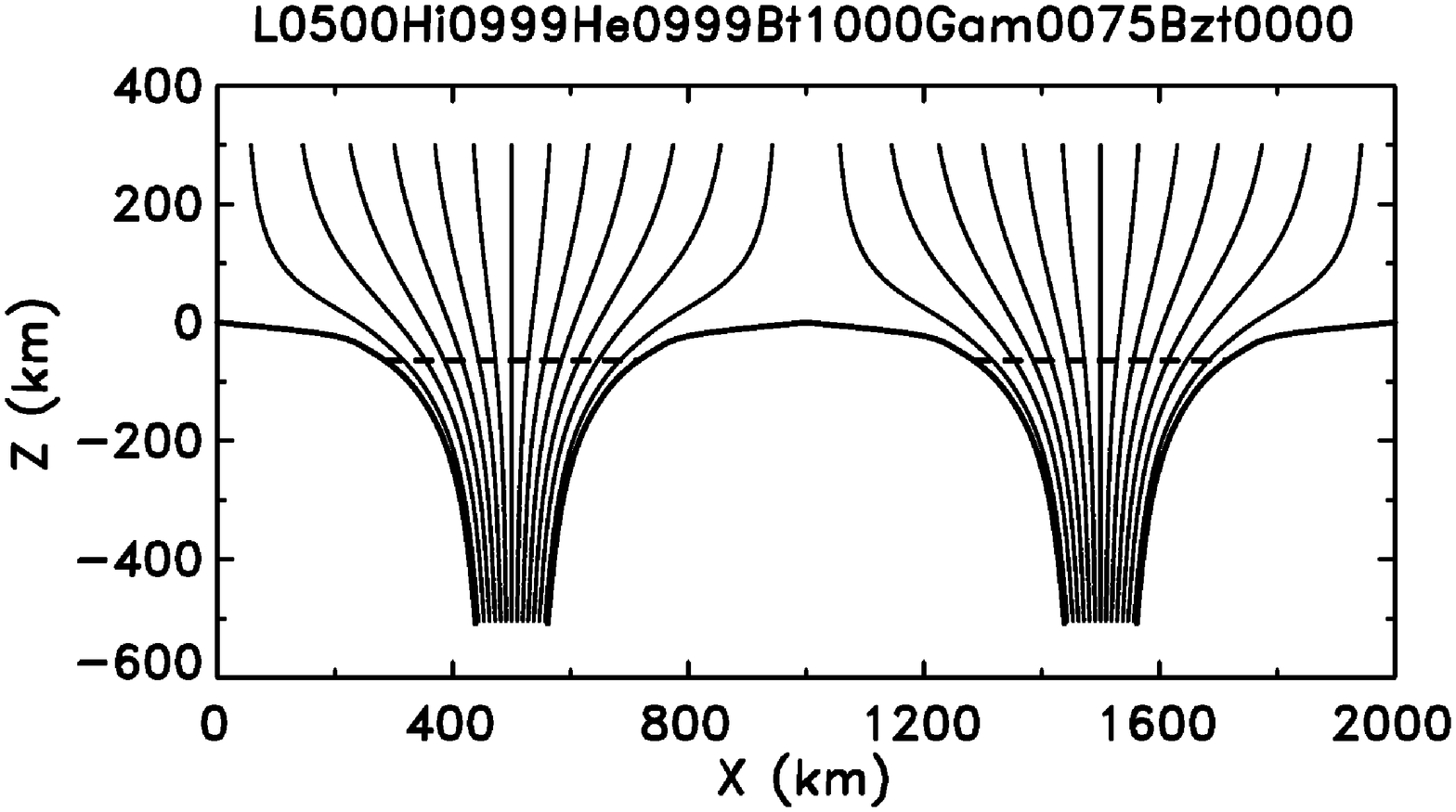}
\hfill
\caption{Magnetostatic convective gap models \citep{2006A&A...460..605S} for the inner (left)
and outer (right) penumbra. The dashed curves show the height at which the gas pressure in the
magnetic component equals the photospheric gas pressure in the field--free component. This serves
as proxy for the $\tau=1$ surface and leads to Wilson depressions on the order of 200~km for
the inner and 50~km for the outer penumbra. Note the similarity of the magnetic field
configuration for the outer penumbra above the `photosphere' in this model and the flux tube
model, shown in Fig\@.\ref{flux_tube}} 
\label{gap}       
\end{figure}

The potential magnetic field configurations associated with two magnetostatic models are 
shown in Fig\@.\ref{gap}. The upper figure shows the calculated field lines for a weak (1000~G) nearly 
horizontal (average inclination 75~deg) magnetic field and the lower figure for a stronger 
(1800~G) and more vertical (average inclination 45~deg). Also shown are the shapes of the gap 
and the height at which the gas pressure between the gaps is equal to that of the field-free 
component at $z=0$ (horizontal dashed lines). This serves as a proxy for the height at which 
the continuum optical depth is equal to unity. As is clear from the figure, the magnetic field
configurations {\em above} the gaps are associated with strong gradients. The simple potential 
field model thus explains large magnetic field inclination variations {\em above} the penumbral 
photosphere without invoking forces in these layers. The associated current sheet is located at 
and below the photosphere, where the gas pressure
is much higher than a few hundred km above the photosphere. This is in contrast to the uncombed
model \citep{1993A&A...275..283S}, where the strong gradient in the magnetic field is a direct 
consequence of a {\em local} perturbation in the form av an embedded flux tube located {\em above} 
the photosphere. The current sheets associated with such flux tubes are difficult to
combine with magnetostatic equilibrium because of the lower gas pressure at these heights 
\citep{2006A&A...447..343S}. Indeed, most of flux tube in the magnetostatic model of 
\citet{2007A&A...471..967B} is buried below the photosphere and only about 130~km protrudes above
the surrounding photosphere. It seems very difficult, if not impossible, to construct similar models for 
flux tubes located entirely above the penumbral photosphere.

The convective gap model thus explains strong magnetic field gradients above the penumbral 
photosphere as a necessary consequence of potential fields. This model eliminates the problems 
of large curvature forces discussed by 
\citet{1992ApJ...398..359S, 1994A&A...283..221S}, constituting a corner-stone
argument in favor of the uncombed penumbra model \citep{1993A&A...275..283S}. The convective
gap model also predicts configurations for the inner and outer penumbra that are quite different. 
In the inner
penumbra, the field is cusp-shaped and associated with a large Wilson depression, large field 
strength fluctuations but relatively small fluctuations in inclination. In the outer penumbra,
the magnetic field is spine-like \citep{1993ApJ...418..928L}, with a small Wilson depression,
and small field strength fluctuations above the photosphere but with large inclination variations.
These qualitative differences between the inner and outer penumbra are in good agreement
with observations \citep{2006A&A...460..605S}.

The overturning convective flow patterns associated with the gaps are predicted to be upward 
in the middle and downflow along the boundaries to the magnetic components 
\citep{2006A&A...460..605S}. Added to this flow pattern is a radially outward (Evershed) flow,
explained by \citet{2008ApJ...677L.149S} on the basis of 3D MHD simulations
\citep{2007ApJ...669.1390H} as being identical to the horizontal component of this convection.
In our model, the {\em dark} cores of the penumbral filaments 
correspond to locations of convective {\em upflows}, in contradiction with what we expect from 
field-free convection. As explained above, the strong fluctuations in field strength across 
the filamentary structures in the inner penumbra lead to a correlation between brightness
and field strength, such that we see deeper in to the hotter gas where the field strength is
high. The mechanism for producing the dark penumbral cores is directly related to the
mechanism that produces bright faculae, proposed initially by \citet{1976SoPh...50..269S} and well
established by numerical simulations \citep{2004ApJ...607L..59K, 2004ApJ...610L.137C,
2005A&A...430..691S} and
high-resolution SST observations \citep{2004SoPh..221...65L}. In the case of faculae,
convection occurs in the surrounding darker and field-free photosphere, whereas convection
is inhibited or strongly suppressed in the brighter and magnetic faculae.

\begin{figure}
\hfill
\center
\includegraphics[clip, width=0.5\textwidth]{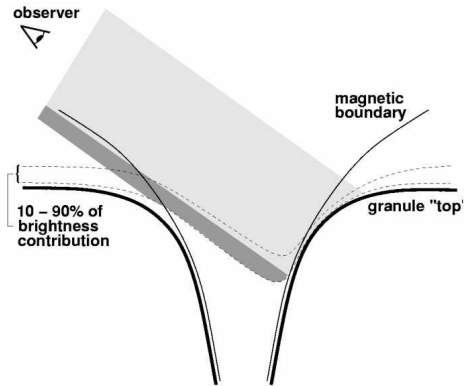}
\hfill
\caption{Schematic drawing explaining the brightness of faculae observed near the limb
\citep{2004ApJ...607L..59K}. The low gas pressure associated with the strong magnetic field
of the flux concentration makes it essentially transparent such that we can see the hot
granular wall of the surrounding denser field--free gas. The optical depth unity surface is 
strongly tilted with respect to the horizontal by the shape of the flux tube magnetic field
such that we see {\em deeper in to} the hot convecting granules close to the limb than at
disk center, explaining the brightness. A similar situation occurs with penumbral filaments
associated with strong variations in field strength (see Fig\@.\ref{gap}). Another similarity
between flux tubes and faculae is that
the radiative cooling of the {\em surrounding} field--free gas through the facula leads to a
convective flow pattern that is downward adjacent to the facula and that this downflow is
observable on the limb--side but hidden from view on the center--side of the facula. In a 
similar way, strong Wilson depressions associated
with narrow penumbral filaments lead to obscuration of the limb--sides of the filaments already
for small heliocentric distance. Convection in such filaments are also affected by radiative
cooling through the surrounding magnetic gas such that we expect upflows in the middle of the
filaments and downflows at their sides.}
\label{facula}       
\end{figure}

The connection between dark--cored filaments and faculae can be carried further, c.f., 
Fig\@. \ref{facula}. The brightness
of faculae near the limb is explained by the $\tau=1$ surface being strongly inclined to the
horizontal at the limbside of faculae
\citep{1976SoPh...50..269S, 2004ApJ...607L..59K, 2004ApJ...610L.137C}. 
Looking through the nearly transparent gas within the strong magnetic field, we see deeper 
into the {\em surrounding}
hot convecting gas close to the limb than at sun center. In a similar way, we see deeper into
the convecting parts of the penumbral filaments when viewing sunspots well away from disk center 
and at +/- 90~degrees from the symmetry line \citep{2006A&A...460..605S}. The convective gap
model leads us to interpret the absence of observational evidence for penumbral convection 
as not only due to lack of spatial resolution: there is also a difficulty of seeing deep enough 
into the filaments to be able to observe the convection for sunspots close to disk center. In 
addition, there is the confusion from the brightness -- field strength correlation already discussed above.

\subsection{Limitations of the convective gap model} 
The magnetostatic gap models discussed above \citep{2006A&A...460..605S} predict a gradual 
transition from cusp-shaped magnetic fields in the inner penumbra to spine-like 
\citep{1993ApJ...418..928L} magnetic field configurations in the outer penumbra, in good
agreement with observations. However, these simple models assume a perfectly field--free gap 
and do not include an energy equation,
nor are the forces associated with the convective flows included. Details of flows and their
interactions with the magnetic field cannot be explained with this simple model.
The explanation of dark
cores relies on opacity effects that are obvious only for the model corresponding to
the inner penumbra. We expect these results to be relatively robust. In the outer penumbra,
predictions about filamentary brightness are more difficult without an energy equation. Furthermore,
observations show strong Evershed flows in the outer penumbra where the magnetic field is
weaker. This combines to making the kinetic energy density $\rho v^2/2$ of comparable
magnitude to the magnetic energy density $B^2/2\mu_0$ such that we expect relatively 
strong effects from the flow on the magnetic field. Whereas the magnetostatic gap models
show good overall agreement with observed properties of penumbral magnetic fields, we cannot 
expect detailed agreement between the gap model and observations also in the outer penumbra. 
Only more accurate models and numerical simulations can provide this. 

\subsection{Support from observations}
The interpretation of light bridges as essentially field--free gaps dividing the umbra of
a sunspot in two parts \citep{1997ApJ...484..900L, 2006A&A...453.1079J} does not seem
controversial. 3D MHD simulations of such structures \citep{2006ASPC..354..353N, 
Heinemann...2006} reproduce
observed dark lanes running along the center of such structures \citep{2004SoPh..221...65L}
and demonstrate that the origin of this dark structure is the same as proposed for the
convective gap model. Of considerable importance is therefore that dark-cored light bridge
structures occasionally show smooth {\em transitions} to dark-cored penumbral filaments
\citep{2006ASPC..358....3L, 2007ASPC..369...71S}, strongly suggesting similar origin.
Upflows in light-bridge dark lanes and the dark cores of penumbral filaments 
\citep{2008ApJ...672..684R} suggest a common interpretation in terms of convection, but with 
evidence for horizontal flows at greater heights also fitting a flux tube interpretation.
Connections of dark-cored penumbral filaments to peripheral umbral dots and dark cores
in light bridges have been reported also by \citet{2007ApJ...669L..57B}.

Several recent papers report evidence for convection in umbral dots \citep{2007ApJ...665L..79B,
2008ApJ...672..684R, 2008ApJ...678L.157R}. Although this provides no direct evidence for
penumbral convection, the direct {\em connection} of peripheral umbral dots to dark-cored 
filaments \citep{2006ASPC..358....3L, 2007A&A...464..763L, 2008ApJ...672..684R} provides 
`circumstantial' evidence for this interpretation. We caution, however, that the umbral dots
observed are much larger than those simulated by \citet{2006ApJ...641L..73S} and in some 
cases resemble granular intrusions.

\begin{figure}
\hfill
\includegraphics[clip, width=\hsize]{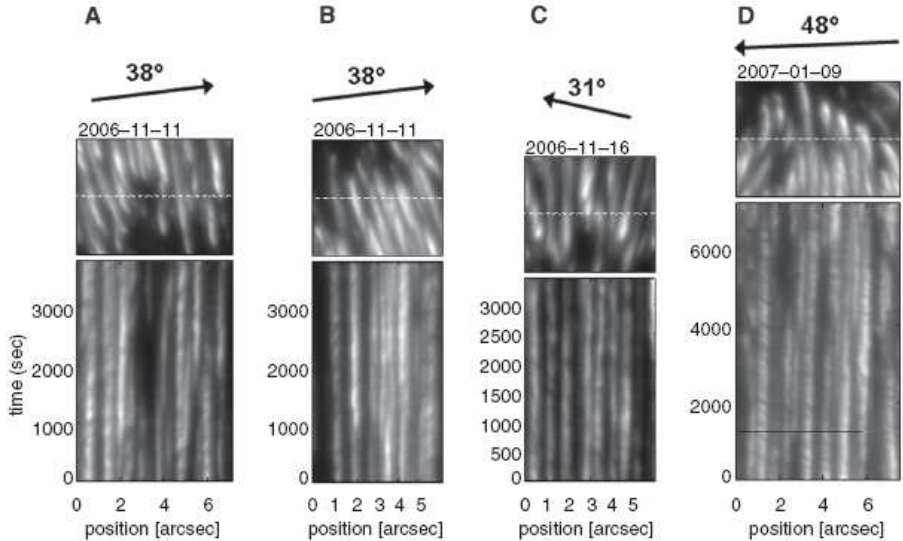}
\hfill
\caption{Examples of space-time slices across filaments in the inner penumbra observed
with SOT/Hinode \citep{2007Sci...318.1597I}. The disk center direction is indicated with
an arrow and labeled with the corresponding heliocentric distance. The upper image of each
pair shows the surroundings of the cut across the filaments (dashed line) analysed in the
space-time slices (lower image).}
\label{twist}       
\end{figure}

In a paper of fundamental importance, \citet{2007Sci...318.1597I} find strong evidence for 
overturning penumbral convection by analyzing continuum images and spectra observed with
SOT on Hinode. Space-time plots created along lines crossing filaments in
the {\em inner} penumbra, at directions +/- 90~deg from the line connecting the center
of the solar disk and the center of the sunspot, displayed twisted ropelike structures. 
The `twisting' motion was consistently in the direction toward sun center for both sides 
of the spot and irrespective of whether the spot was East or West of the meridian. 
(c.f, Fig\@.\ref{twist}). The apparent twist observed is therefore neither an actual 
twist nor a helical motion of individual filaments but must be a viewing angle effect.
The interpretation \citep{2007Sci...318.1597I} is of upflows of overturning convection, 
viewed from the side. With the limb-side part of these filaments hidden from view, such
flows will always appear to be in the direction of sun center direction for spots observed
away from disk center. This is in perfect agreement with the predictions of the magnetostatic 
gap model for the {\em inner} penumbra \citep{2006A&A...460..605S}. Here the Wilson depression
between the gaps is predicted to be so large that the limbside part of the filaments is 
invisible already for disk center distances in the range 21--35~deg. The spot observed by 
Ichimoto et al. was located at a disk center distance of 31--48~deg. The flow pattern observed is 
consistent with that predicted by \citet{2006A&A...460..605S}.

Based on high-resolution SST data, \citet{2008A&A...488L..17Z} recently inferred similar 
evidence of convective flows for a sunspot located 40~deg from disk center. Surprisingly, 
this is interpreted in terms of convective rolls \citep{1961ApJ...134..289D}. However, the 
horizontal flow toward the {\em center} of the filament at the {\em bottom} of such a roll, needed 
to verify its existence, is unobservable. Furthermore, the life times of penumbral filaments 
are on the order of 1 hour or more \citep{2007A&A...464..763L}. To sustain the radiative output 
over such a long time, 
the convective upflow must persist to depths much larger than a few hundred km. Finally, the 
observations of \citet{2007Sci...318.1597I} were made in the {\em inner} penumbra, where the
magnetic field has a strong vertical component, whereas 
roll-like convection is expected to be primarily associated with more horizontal magnetic 
field.

\subsection{Support from 3D MHD simulations}
In contrast to what is to what is the case for faculae \citep{2004ApJ...607L..59K, 
2004ApJ...610L.137C} and umbral dots \citep{2006ApJ...641L..73S}, realistic simulations of
entire sunspots have not yet been feasible. This is partly due to the difficulties
of thermally relaxing such a deep structure and maintaining its stability but mostly due to 
the huge range of scales associated with a fully developed sunspot. The first attempts to 
carry out 3D MHD simulations with radiative energy transfer of fine structure in a small 
sunspot were carried out by \citet{2007ApJ...669.1390H} and further discussed by 
\citet{2008ApJ...677L.149S}. Recently, simulations using a similar setup and grid separation,
but with a much larger computational box, were carried out by \citet{2008arXiv0808.3294R}. 
A synthetic continuum image calculated from these simulations is shown in Fig\@. \ref{3DMHD}.
The approach taken in both simulations is to reduce the computational effort by using a rectangular 
computational box containing only a small `azimuthal' slice of a sunspot. 
The overall results of these simulations obtained with two independent codes are quite similar,
although differing strongly in the length of the penumbral filaments:

\begin{itemize}
\item The origin of filamentary structures is overturning convection
      and the dark cores are caused by a locally elevated $\tau=1$ surface, supporting the
      convective gap model \citep{2006A&A...447..343S}.
\item The convection occurs in deep gaps, up to about 2~Mm \citep{2008arXiv0808.3294R}
      with strongly reduced field strength.
\item The simulations show horizontal outflows, similar to Evershed flows but with smaller velocities,
      peaking near optical depth unity and associated with locally strongly inclined fields.
\item The bright heads of the penumbra filaments show inward propagation and strong upflows 
\item The simulations show moving magnetic features (MMF's) and moat flow in the surrounding
      photosphere 
\end{itemize}

For overturning convection to be efficient, the gas needs to stay near the surface for a significant
amount of time in order to give it time to cool radiatively. At the same time, it needs to move
away from its upflow point in order to allow more gas to flow up. {\em Horizontal} flows are 
thus essential components of overturning convection.
The Evershed flow is identified as being {\em identical} to the horizontal flow component of this
penumbral convection  \citep{2008ApJ...677L.149S}. 

\begin{figure}
\hfill
\includegraphics[clip, width=\hsize]{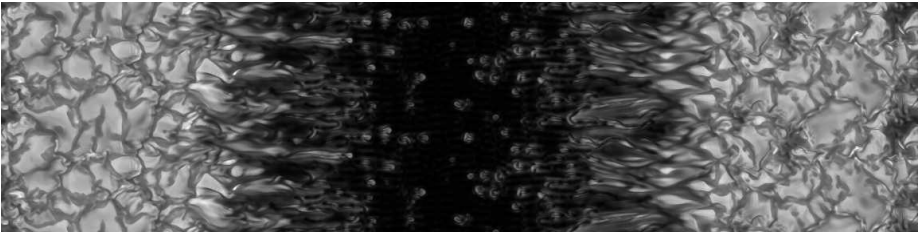}
\hfill
\caption{Synthetic continuum image calculated from 3D MHD simulations, showing a sunspot and its 
surrounding photosphere \citep{2008arXiv0808.3294R}. In the umbral part of the spot is seen
umbral dots with dark cores, in the penumbral part filaments that reach lengths of up to 2--3~Mm.}
\label{3DMHD}       
\end{figure}

The simulations reproduce fundamental properties of observed penumbrae.
This gives us confidence in concluding that the simulations constitute good representations 
of penumbral dynamics and energy balance. However, neither of these penumbra simulations show 
nearly horizontal magnetic fields in the outer penumbra, as observed, so something is missing
in the simulations. Other features that do not appear realistic are elongated structures in
the outer penumbra that appear intermediate to granules and filaments and a less distinct boundary
to the umbra than observed. A complex question concerns the field strengths in the gaps. In
the simulations these are on the order of 700--1000~G. In the simulations of 
\citet{2007ApJ...669.1390H}, this was interpreted as a consequence of numerical
diffusivities, allowing overturning convective flows to cross field lines. Similar processes
take place in the simulations of Rempel et al. (Rempel, private communication). The
existence of strong magnetic fields associated with overturning convection in penumbrae thus 
rely on turbulent magnetic diffusion at small scales. In contrast, umbral dots
in the simulations of \citet{2008arXiv0808.3294R} as well as those of \citet{2006ApJ...641L..73S} 
appear as nearly field--free plumes near the surface.

The simulations of Rempel et al. show some indication of flow patterns that are
reminiscent of roll convection \citep{1961ApJ...134..289D}, but this interpretation is uncertain.
The convective energy transport is mainly provided by {\em deep} upflow 
and downflow plumes. An intriguing result of both simulations is
that the convective gaps do not extend to the field-free atmosphere below the sunspot,
suggesting that sunspots radiate energy that is contained within the magnetic field 
initially. A difference between the simulations of Rempel et al. and Heinemann et al. is that
the former simulations show radial {\em inflows} adjacent to the outflows near the surface
whereas the latter simulations show evidence of such return flows only deeper below the
surface.

A perhaps relevant result for understanding penumbral magnetic fields comes from simulations 
of quiet sun emerging flux. In these simulations, the expulsion of magnetic flux takes place 
in the horizontal direction by horizontal flows but also in the vertical direction by 
overshooting convection \citep{2008ApJ...680L..85S}. The weak horizontal fields of such emerging 
flux are expelled to heights of about 500~km, where they are difficult to observe in
photospheric lines. Such expelled magnetic fields are limited to a height at 
which the gas pressure below the expelled magnetic field is roughly equal to $B^2/2\mu_0$. 
If a similar expulsion mechanism
operates in the penumbra, where the field strength is much higher, any expelled horizontal 
field must be located much closer to the penumbral photosphere than for the emerging flux
discussed by \citet{2008ApJ...680L..85S}. Such magnetized gas would then show strong polarization 
signatures reminiscent of flux tubes. 

\subsection{Connections to flux tube models}

In spite of its 1D representation and the failure to confirm interchange convection, the moving
tube model \citep{1998ApJ...493L.121S, 1998A&A...337..897S, 2002AN....323..303S} has connections 
to convective processes. The upflow {\em within} the flux tube is driven by the superadiabatic
stratification of the external atmosphere, similar to that of a field-free convective upflow.
The flux tube evolution is similarly driven by the superadiabatic stratification of the external
atmosphere, even though the complete cycle of inward/outward movement and heating/cooling of 
the flux tubes does not take place in the simulations: a flux tube initially located along the 
magnetopause moves toward the umbra and remains there. These 1D simulations cannot show 
convective flow patterns. However, the `sea serpent' behavior \citep{2002AN....323..303S}, 
with upflows and downflows along the length of the tube, can be interpreted as a 1D 
representation of 3D convection.  In the 3D penumbra
simulations \citep{2007ApJ...669.1390H, 2008arXiv0808.3294R}, the upflows occur at the centers 
of the gaps and the downflows on either side of the upflows, displaced both the in the azimuthal 
and radially outward directions. In the moving tube model, such upflows and downflows can 
only be spatially separated in the radial direction. 
Both models have in common a convective upflow and radiative cooling driving an outflow away 
from the center of the spot. In both moving tube and 3D penumbra simulations, the 
outflow peaks in a thin layer near $\tau=1$, which is where the gas cools most efficiently
\citep{2008ApJ...677L.149S}.

In the convective gap model, the strong magnetic field gradients above the gap are explained
as a perturbation of a nearly potential magnetic field above the penumbra, introduced by the 
nearly field--free gap \citep{2006A&A...447..343S, 2006A&A...460..605S}. This 
leads to a magnetic field that is cusp--shaped in the inner penumbra and locally nearly horizontal 
in the outer penumbra. Adding a weak horizontal magnetic inside the gap will not change this
configuration significantly. Such a configuration is in its {\em upper} 
parts quite similar to a flux tube but deeper down, these two types of structures are very 
different. 

Support for the existence of embedded flux tubes based on a magnetostatic model 
is claimed from calculations of net circular polarization (NCP),
\citep{2007ApJ...666L.133B}. However, the $\tau=1$ surface of these models intersect the
symmetry axis of the flux tube well above the center of the flux tube and the $\tau=0.1$ surface
(typical of the line formation height) cuts through the top of the flux tube (c.f., 
Fig\@. \ref{flux_tube}). These calculations clearly are sensitive only to the upper part of the 
flux tube, where its magnetic field is similar to that of the convective gap model. 

A similar ambiguity concerns the origin of the dark cores of penumbral filaments, explained 
by convective gap models \citep{2006A&A...447..343S, 2006A&A...460..605S}.
Also flux tube models with {\em weaker} field in the flux tube than in the surroundings 
produce opacity effects resulting in dark--cored structures \citep{2008A&A...488..749R}.
This is interpreted as support for flux tube models by the authors. However, 
also for this model, only the top of the flux tube is visible above $\tau=1$, so this 
configuration is similar to a convective gap model in its observable parts. 

Interpretations of highly resolved Stokes spectra SOT/Hinode show wrapping around structures 
that can similarly be interpreted either as flux tubes or as convecting gaps 
\citep{2008A&A...481L..13B}. These and other recent high-resolution Stokes data do not 
provide evidence for flow channels and flux tubes {\em elevated} above the photosphere, as 
discussed in some papers, e.g., \citet{1993A&A...275..283S, 1995A&A...298..260R, 
2006A&A...450..383B}. Based on observations of penumbra magnetic fields, interpretations in 
terms of convective gaps or flux tubes partly buried below the $\tau=1$ surface thus are 
inherently ambiguous. This also serves as a reminder of the difficulties of interpreting 
(inadequately resolved) observations in terms of unknown underlying physics.

While this ambiguity, in our opinion, undermines arguments for the very existence of embedded flux 
tubes, it primarily suggests that magnetic field measurements are not likely to show a clear
distinction between flux tube and convective gap models, at least in the outer parts of the
penumbra. The distinction between the models may need to be based primarily on measurements 
of the velocity field, which is horizontal and along a flux tube but with added vertical and 
azimuthal flow components in the convective gap models. It is this 
diagnostics that so far provides the strongest observational evidence in favor of the 
convecting gap model \citep{2007Sci...318.1597I}. The observational evidence accumulated so far
is however too scarce to be conclusive. \citet{2008A&A...488L..17Z} also reported evidence for
convective upflows in penumbral filaments, but clearly observations of the entire sequence 
of upflows, horizontal flows and downflows are needed to fully reveal the nature of convection 
in penumbrae.

A highly controversial issue is whether the penumbral convection is essentially field--free or 
associated with kG strengths. \citet{2007ApJ...668L..91B}, based on SOT/Hinode data and 
Milne--Eddington (ME) inversions, found only small variations in field strength across 
dark--cored penumbral filaments. \citet{2008arXiv0806.1638S}, based on SST observations and 
ME inversions, demonstrated that an improvement of the spatial resolution from 0.3" to 0.15" 
increases measured field strength variations over dark cores by approximately a factor 
of two. \citet{2008A&A...488L..17Z}, based on SST data and ME inversions, found locally weaker
fields by approximately a factor two, associated with convecting filaments. 
\citet{2007PASJ...59S.601J} using SOT/Hinode data and inversions allowing for gradients 
and a Gaussian perturbation in the magnetic field to represent flux tubes, concluded that 
the field strength is reduced by only 600~G at the centers of bright filaments (where dark 
cores should be located) in the inner penumbra and furthermore that this reduction in field 
strength occurs only close to the photosphere and disappears already at $\log \tau =-0.5$. 
Based on observations and inversions, there is so far no support for the assumption that these 
gaps are nearly field--free. This conclusion refers to observational data from the layers 
{\em above} the photosphere whereas the nature of penumbral convection in deeper
layers can ultimately only be determined from 3D MHD simulations. 

\section{Conclusions}
We believe that there is now strong evidence to support the conclusion that penumbral
fine structure should be interpreted as the result of overturning convection, as proposed
by \citet{2006A&A...447..343S}. Evidence for this conclusion comes from recent 
SOT/Hinode and SST observations \citep{2007Sci...318.1597I, 2008A&A...488L..17Z}, 
showing {\em vertical} flows of the right magnitude to explain the penumbral radiative
heat flux. Recent numerical 3D MHD simulations \citep{2007ApJ...669.1390H, 2008arXiv0808.3294R}
reproduce fundamental properties of observed penumbrae and confirm the convective origin of 
penumbral filaments. The simulations show that the nature of this convection takes place
in gaps with up to 2~Mm depth and that any roll-like convection \citep{1961ApJ...134..289D,
2004ARA&A..42..517T, 2008arXiv0808.3294R}, if present, is of small importance 
\citep{2008arXiv0808.3294R}. The Evershed flow is
interpreted to be identical to the horizontal flow component of this convection 
\citep{2008ApJ...677L.149S}. Such horizontal flows are necessary in order to cool
hot upflows by radiation.

Neither observations nor simulations lead to the conclusion that this convection is nearly 
field-free, as suggested \citep{2006A&A...447..343S}. However, inferred field strengths
from spectropolarimetric data are obviously limited to 
layers above the photosphere, whereas simulations rely on numerical diffusivities
to prevent instabilities at scales corresponding to the grid separation. Other uncertainties
relate to the outer parts of penumbrae where observations show nearly horizontal field and
even field lines dipping down into the photosphere \citep{2008arXiv0806.4454B}. Simulations 
do not show structures
of this type. It appears likely that downward pumping of magnetic field by convection outside
the sunspot plays a role in the outer penumbra, as proposed earlier 
\citep{2002AN....323..383T}, but we disagree strongly with the conclusion that this explains the 
origin of the filamentary structure of the penumbra. Downward pumping by the convection
{\em inside} the penumbra also must take place and this probably explains why observations 
\citep{1997Natur.389...47W} show evidence of return flux well inside the outer penumbral boundary 
\citep{2008ApJ...677L.149S}.

Penumbral filaments have been successfully interpreted in terms of embedded flux tubes during a 
period of 15 years. While we conclude that this interpretation is misleading in terms
of underlying physics, there
are several reasons why this model has been so successful. We have shown that the opening of
radially aligned gaps with nearly field-free convecting gas leads to a magnetic field that
is much more horizontal over the gaps, giving the illusion of a flux tube 
\citep{2006A&A...447..343S}. There are also other arguments for expecting nearly horizontal fields
in the penumbral atmosphere: Horizontal cooling 
flows are most efficient near optical depth unity and if this gas is magnetized, it will aid in 
producing nearly horizontal magnetic fields. Also, emerging flux
simulations relevant to the quiet sun suggests that convection can lead to flux expulsion in the
vertical direction in addition to the horizontal direction and that this explains quiet sun
horizontal magnetic fields above the photosphere. Clear evidence of such `vertical' flux expulsion 
is however not seen in the penumbral part of the sunspot simulated by \citet{2008arXiv0808.3294R}. 

We emphasize the connections of the moving tube model to convective processes and to the 
radiative cooling of such flows near the photosphere in both types of models. The 
similarity of the magnetic field above convecting gaps and flux tubes add to the difficulties 
of correctly interpreting observations and distinguishing between models.

In the embedded flux tube models, the Evershed flow is at center stage and the mechanism for
heating the penumbra remains obscure. The new view of penumbral fine structure as caused by
overturning convection implies that the main driver of penumbra fine structure is the energy 
flux below the surface and that the Evershed flow is `only' a consequence of this convection 
\citep{2008ApJ...677L.149S}.

We believe that siphon flow models \citep{1997Natur.390..485M} are of little relevance for 
understanding penumbrae. These
are linked to the idea that there are two distinct families of field lines: those associated
with dark filaments and Evershed flows and those associated with bright filaments connecting 
to distant magnetic regions \citep{2004ApJ...600.1073W, 2004ARA&A..42..517T}. As far as we know, 
there is no observational support for this
`static' picture of penumbral magnetic fields. Observations suggest life times for 
penumbral filaments on the order of one hour associated with flow channels opening
and closing continuously \citep{2006ApJ...646..593R}. 3D MHD Simulations \citep{2008arXiv0808.3294R,
2007ApJ...669.1390H} confirm the transient nature
of azimuthal variations in field strength and inclination. Theoretical arguments 
and models \citep{2006A&A...447..343S, 2006A&A...460..605S} as well as simulations
clearly lead to the conclusion that the large variations in inclination across
filaments are  {\em local} perturbations, caused by {\em penumbral} convection and vanishing 
a few hundred km above the penumbral photosphere. 

As regards further progress in this rapidly evolving field, we expect that even more realistic 
3D MHD simulations in the near future will further improve our understanding of penumbrae, in 
particular as regards their outermost parts. Observed unpolarized and polarized spectra at the 
highest possible spatial resolution are needed. Of particular importance is 
such spectra giving information about the layer immediately above the photosphere. Emphasis 
should be given to analyzing data at +/- 90~deg from the symmetry axis of 
sunspots located away from disk center, as was done by \citet{2007Sci...318.1597I}. This
is in part to allow analysis of flows perpendicular to the radial direction of the filaments,
but also in order to see as deep into these structures as possible. Analysis of such data
need to account for the pronounced 3D nature of these filaments, caused by strong azimuthal
variations in the Wilson depression, as well as strong LOS variations in the magnetic field
and flow velocity. A dilemma here is that the use of inversion techniques with many nodes
along the LOS raises questions of uniqueness and difficulties in comparing the results of 
such inversions with simulations. Existing 3D MHD simulation data allow 
inversion techniques to be tested with synthetic Stokes spectra from penumbral atmospheres, 
as done already with simulations of small-scale flux concentrations outside sunspots 
\citep{2007MmSAI..78..166K, 2007ApJ...662L..31O}. The effect of assuming e.g. hydrostatic
equilibrium can be evaluated quantitatively.

Presently used inversion techniques process 
polarized spectra pixel by pixel without constraining, for example, the magnetic field to be 
divergence-free. The requirement of divergence--free magnetic fields is crucial in forcing
field lines to bend around convecting gaps (and flux tubes), leading to strong gradients
in field strength and inclination. With spectropolarimetric observations approaching a spatial
resolution of 100~km \citep{2008arXiv0806.1638S}, which is similar to the equivalent LOS 
resolution achieved with inversion techniques using a small number of nodes, it is reasonable
to enforce magnetic fields constrained by div({\bf B})=0. Stray-light corrections are with
most inversion techniques implemented in an ad-hoc manner pixel by pixel whereas 
a physical stray-light implementation would employ a point spread function that does not 
vary, or varies slowly, across the FOV.
Micro-- and macro-turbulence parameters are used as fudge parameters to compensate spatial
smearing of unresolved structures. With improved spatial resolution, modeled LOS velocity 
gradients should eliminate the need for such parameters
in the inversions. We expect future inversion techniques to develop as `global' 
techniques, in the sense of fitting model parameters for a large number of connected pixels
simultaneously. This will allow constraints, such as div({\bf B})=0, and physical straylight 
models to be incorporated in a consistent manner to further enhance the usefulness of 
inversion techniques for inferring the physical state of the atmospheres above sunspots and 
other magnetic structures. 

\begin{acknowledgements}
The author gratefully acknowledges discussions with Henk Spruit, {\AA}ke Nordlund, Rolf
Schlichenmaier, Matthias Rempel, Manfred Sch{\"u}ssler and Oskar Steiner. The author also thanks
Rolf Schlichenmaier for valuable comments on the manuscript.
\end{acknowledgements}

\bibliographystyle{aa}


\end{document}